\documentclass[twocolumn,showpacs,aps,prl,superscriptaddress,amssymb,floatfix,preprintnumbers]{revtex4}
\usepackage{graphicx,rotating}
\usepackage{dcolumn}
\usepackage{epsfig}
\usepackage{multirow}

\input babarsym

\long\def\inst#1{\par\nobreak\kern 4pt\nobreak
    {\it #1}\par\vskip 10pt plus 3pt minus 3pt}

\begin{document}
\preprint{ \babar-PUB-06/042}
\preprint{SLAC-PUB-12113}
\preprint{hep-ex/0609027}

\title{
\Large \bf \boldmath
 Observation of \btppk\ and evidence for \bztppkz\ below \etac\ threshold}

%
\author{B.~Aubert}
\author{M.~Bona}
\author{D.~Boutigny}
\author{F.~Couderc}
\author{Y.~Karyotakis}
\author{J.~P.~Lees}
\author{V.~Poireau}
\author{V.~Tisserand}
\author{A.~Zghiche}
\affiliation{Laboratoire de Physique des Particules, IN2P3/CNRS et Universit\'e de Savoie,
 F-74941 Annecy-Le-Vieux, France }
\author{E.~Grauges}
\affiliation{Universitat de Barcelona, Facultat de Fisica, Departament ECM, E-08028 Barcelona, Spain }
% ...
\collaboration{The \babar\ Collaboration}
\noaffiliation

\date{November 23, 2006}

\begin{abstract}

 We report measurements of the decays \btppk\ and \bztppkz\ using a sample of 231 million $B\Bbar$ pairs collected with the \babar\ detector at the PEP-II asymmetric-energy \B\ Factory at the Stanford Linear Accelerator Center. The branching fractions are measured to be \BRbtppk\ = ($7.5 \pm 1.0$ \stat $\pm$ 0.7 \syst) $\times$ $10^{-6}$  and \BRbztppkz\ = ($4.1^{+1.7}_{-1.4}$ \stat $\pm$ 0.4 \syst) $\times$ $10^{-6}$ for a $\phi\phi$ invariant mass below 2.85 \gevcc. 
\end{abstract}
\maketitle

\setcounter{footnote}{0}
We report an observation of the decay \btppk\ and evidence for \bztppkz\ along with their corresponding branching fractions. The decay modes studied involve a flavor-changing neutral current $b \to s\bar{s}s$ transition. These charmless transitions can interfere with the $\b \to \ccbar\s$ process \B \to \etac \kaon, \etac \to \pp\ and lead to direct $\emph{CP}$ violation~\cite{Hazumi}; the $\emph{CP}$ asymmetry expected in the Standard Model (SM) is zero, so a non-zero $\emph{CP}$ asymmetry would be a sign of new physics. Furthermore, an analysis of time-dependent \CP violation in $\Bz \to \phi \phi \Kz$ would be sensitive to physics beyond the Standard Model and complementary to measurements in the other decays that are dominated by the $\b \to \s\sbar\s$ transition. In the SM, the partial decay widths for these decays are expected to be equal due to the suppression of $\Delta I$ = 1 transitions in the electroweak Hamiltonian~\cite{F&M}. Additional interest in these final states arises from the possibility of glueball production with subsequent decays to \pp~\cite{gball}. 

 We study the charmless decays $B\to \phi\phi K$ by working below the charm production threshold (\mpp\ $< 2.85$ \gevcc) to avoid the region dominated by the \etac\ resonance. Theoretical estimates of these branching fractions are in the range (1.3 -- 4.2) $\times~10^{-6}$~[4,~5] within the above kinematic region. The Belle Collaboration has previously reported evidence for the decay \btppk\  with a branching fraction of $2.6^{+1.1}_{-0.9}$ \stat\ $\pm$ 0.3 \syst\ $\times$ $10^{-6}$ for \mpp\ $<$ 2.85 \gevcc\ \cite{belle}; no measurement of the branching fraction for \bztppkz\ has previously been reported. Throughout this paper, for any given mode, the corresponding charge-conjugate mode is also implied.
 
  The data used in this analysis were collected with the \babar\ detector~\cite{ref:babar} at the \pep2\ asymmetric \epem\ storage ring. These data represent an integrated luminosity of 209.1 \invfb\ collected at a center-of-mass (CM) energy $\sqrt{s}$ = 10.58 \gev, near the peak of the \FourS\ resonance, plus 21.6 \invfb\ collected at a CM energy approximately 40 \mev\ below the \FourS. These are referred to as the on-resonance and off-resonance data samples, respectively.

  Charged particles from the \epem\ interactions are detected and their momenta measured by a five-layer, double-sided silicon vertex tracker (SVT) and a 40-layer drift chamber (DCH) with a helium-based gas mixture, placed in a 1.5-T  uniform magnetic field produced by a superconducting magnet. The charged particles are identified using likelihood ratios calculated from the ionization energy loss (\dedx) measurements in the SVT and DCH, and from the observed pattern of Cherenkov light in an internally reflecting ring-imaging detector. A  \kaon/$\pi$ separation of better than four standard deviations ($\sigma$) is achieved for momenta below 3 \gevc, smoothly decreasing to 2.5 $\sigma$ at the highest momenta present in the \B-decay final states. Photons and electrons are identified as isolated electromagnetic showers in a CsI(Tl) electromagnetic calorimeter. The detector response is simulated with the GEANT4 \cite{G4} program. 

\begin{table*}[t]
\caption{ Fitted signal yield, detection efficiency $\epsilon$$(\%)$ including tracking, PID efficiency and fit bias correction, daughter branching fraction product $\prod$$B_{i}$~\cite{PDG}, significance $S$ ($\sigma$), measured branching fraction $\cal{B}$ with statistical and systematic uncertainties for each decay mode. These branching fractions are for $m_{\pp}<2.85$ \gevcc. The first uncertainty is statistical, the second systematic.}

\begin{center}
\tabcolsep=5mm
\begin{tabular}{ c c c c c c } \hline\hline
Mode& Signal Yield& $\epsilon$$(\%)$ & $\prod$$B_{i}$ ($\%$)& $S$($\sigma$) & $\cal{B}$$(10^{-6})$ \hspace{30mm}\\ \hline
\btppk     & $64\pm 9$  &15.3    & 24.2 & 12.9 & $7.5\pm 1.0 \pm 0.7$ \hspace{30mm}\\ 
 \bztppkz  & $10^{+4.1}_{-3.4}$ & 12.6   & 8.3  & 4.2  & $4.1^{+1.7}_{-1.4} \pm 0.4$\\ \hline\hline
\end{tabular}
\end{center}
\label{result}
\end{table*}

   The \B-meson daughter candidates are reconstructed through their decays \phikk\ and \kspipi. For \phikk, we require one charged track to be consistent with the kaon hypothesis, the other to be inconsistent with the pion hypothesis, and the invariant mass to satisfy 1000 $< m_{\KpKm} <$ 1050 \mevcc. The variable $m_{\KpKm}$ will be later used in the fit. The \KS\ candidates are formed from pairs of oppositely charged tracks consistent with the pion hypothesis, with a vertex $\chi^{2}$ probability greater than 0.001 and a reconstructed decay length greater than 2 mm. We require the invariant mass of the two pions to satisfy 486 $< m_{\pippim} <$ 510 \mevcc.

 We reconstruct a \B-meson candidate by combining a \Kp\ or \KS\ with two \mphi\ candidates. From the kinematics of the \FourS\ decays, we determine the energy-substituted mass \mes\ = $(\left(\sqrt{s}/2\right)^{2} - {\pcm}^{2})^{1/2}$ and the energy difference $\DeltaE = E^{*}_{B} - \sqrt{s}/2$, where \pcm\ and $E^{*}_{B}$ are the reconstructed 3-momentum and energy of the \B\ meson calculated in the CM frame, respectively, and $\sqrt{s}$ is the \epem collision energy in the CM. For signal decays, the \mes\ distribution peaks near the nominal mass of the \B\ meson and \de\ peaks at zero. The \de\ (\mes) resolution is about 20 \mev\ (3.0 \mevcc). We require $|\de|\leq 0.2$ \gev, $ \mes > 5.2$ \gevcc, and the invariant mass of the pair of \mphi\ meson candidates to be less than 2.85 \gevcc. The average number of reconstructed \B\ candidates per event is 1.06 (1.05) for \btppk\ (\bztppkz). In events with multiple candidates we arbitrarily select one candidate to avoid a potential bias in the shape of the variables used in the selection

   Backgrounds arise primarily from random combinations of tracks in the continuum \epem\ $\to q\bar{q}$ ($q= u,d,s,c$) events. Because of the jet-like topology, in contrast to the nearly isotropic distribution of final particles from the process \FourS \to \bbbar, 

 the continuum background can be significantly reduced by an appropriate choice of variables describing the event shape. Discrimination between signal and continuum events is obtained using a Fisher discriminant $\cal{F}$. The variable $\cal{F}$ combines eleven event-shape variables defined in the CM frame~\cite{fisher}: the polar angles of the \B\ momentum vector and the \B\ candidate thrust axis with respect to the beam axis, and the scalar sum of the momenta of charged particles and photons (excluding particles from the \B\ candidate) in nine $10^{\circ}$ polar-angle intervals coaxial with the \B-candidate thrust axis. 

    We use Monte Carlo (MC) simulation for an initial estimate of the residual \BB\ background and to identify the decays that may survive the candidate selection and have characteristics similar to the signal. We find that the contributions from the multi-kaon decays, $B^{+/0} \to \mphi \Kp \Km K^{+/0}$ and $B^{+/0} \to \Kp \Km \Kp \Km K^{+/0}$, are negligible after selecting events with two $\phi$ meson candidates.

   We obtain the signal yields from an unbinned extended maximum-likelihood fit. The variables used in the fit are \de, \mes, the invariant masses of two \mphi\ meson candidates, and $\cal{F}$. The likelihood function has two categories of probability-density functions (PDF), one for signal and the other for the continuum background. The likelihood function is defined as
\begin{equation}
{\cal L} =  e^{-\left(\sum n_j\right)} \prod_{i=1}^N \left[\sum_{j=1}^2 n_j  {\cal P}_j({\bf x}_i)\right] ,
\end{equation}
where  $N$ is the number of candidates,  $n_j$ is the number of events in
category $j$, and     ${\cal P}_j ({\bf x}_i)$ is the corresponding PDF, evaluated with the observables ${\bf x}_i$ of the $i$th event. Since correlations among the observables are small, we take each ${\cal P}$ as the product of the PDFs for the separate variables. Possible systematic effects arising from correlations are discussed later. 
 
   We determine the signal PDF parameters from MC simulated data. We generate signal MC assuming that the \B\ meson decays isotropically to \ppk, using three-body phase space. The signal PDF distributions are parametrized using a single Gaussian function for \mes, a sum of two Gaussian functions with the same mean for \de, a sum of an asymmetric Gaussian function with a different width below and above its maximum, and a single Gaussian for $\cal{F}$. The \mphi\ candidate mass distributions are parametrized using a relativistic Breit--Wigner distribution convolved with a Gaussian resolution function. Control samples (e.g., $B\to D(K\pi\pi)\pi$) are used to verify the resolutions obtained from signal MC. The signal PDFs are obtained using correctly reconstructed \bppk\ decays from MC simulated data.

  The background PDFs are determined using \mes\  and \de\ sideband data ($5.20 <\mes <5.26$ \gevcc, $0.1<|\de|<0.2$ \gev). We use a first-order polynomial for \de, an empirical phase-space function~\cite{Argus} for \mes, and an asymmetric Gaussian function  for $\cal{F}$. Since the background includes both resonant and non-resonant \KpKm combinations, the \mphi-candidate mass distributions are parametrized as the sum of the \mphi\ lineshape (as described above) and a first-order polynomial. The parameters allowed to vary in the fit are the signal and background yields and all the background PDF parameters except the \mphi\ mass and width. The signal yield from a fit performed on off-resonance data was consistent with zero, as expected.

 Before applying the fitting procedure to the data we evaluate the possible signal-yield bias from neglecting small residual correlations between discriminating variables in the signal PDFs. The bias is determined from ensembles of mock experiments obtained from samples of signal MC events combined with $\q\bar{q}$ background events generated from the PDFs. We find a bias of $7\%$ ($10\%$) for \btppk\ (\bztppkz). We correct the signal-detection efficiency for this fit bias.

   We compute the branching fractions from the fitted signal-event yields, detection efficiencies, daughter branching fractions, and the number of produced \B-meson pairs. In Table~\ref{result}, we show the fitted signal yield, the detection efficiencies, the products of daughter branching fractions for each decay mode, the significances $S(\sigma)$, and the measured branching fractions. We assume equal decay rates of the \FourS\ to \BpBm\ and \BzBzb. The statistical uncertainties in the signal yields are taken as the change in the central value when the quantity $-2\ln\cal{L}$ increases by one unit from its minimum value. The significance is taken as the square root of the difference between the value of --2ln$\cal{L}$ (with systematic uncertainties included) for zero signal and its value at the minimum.

\begin{figure}[t]
\begin{center}
\includegraphics[width=3.5in]{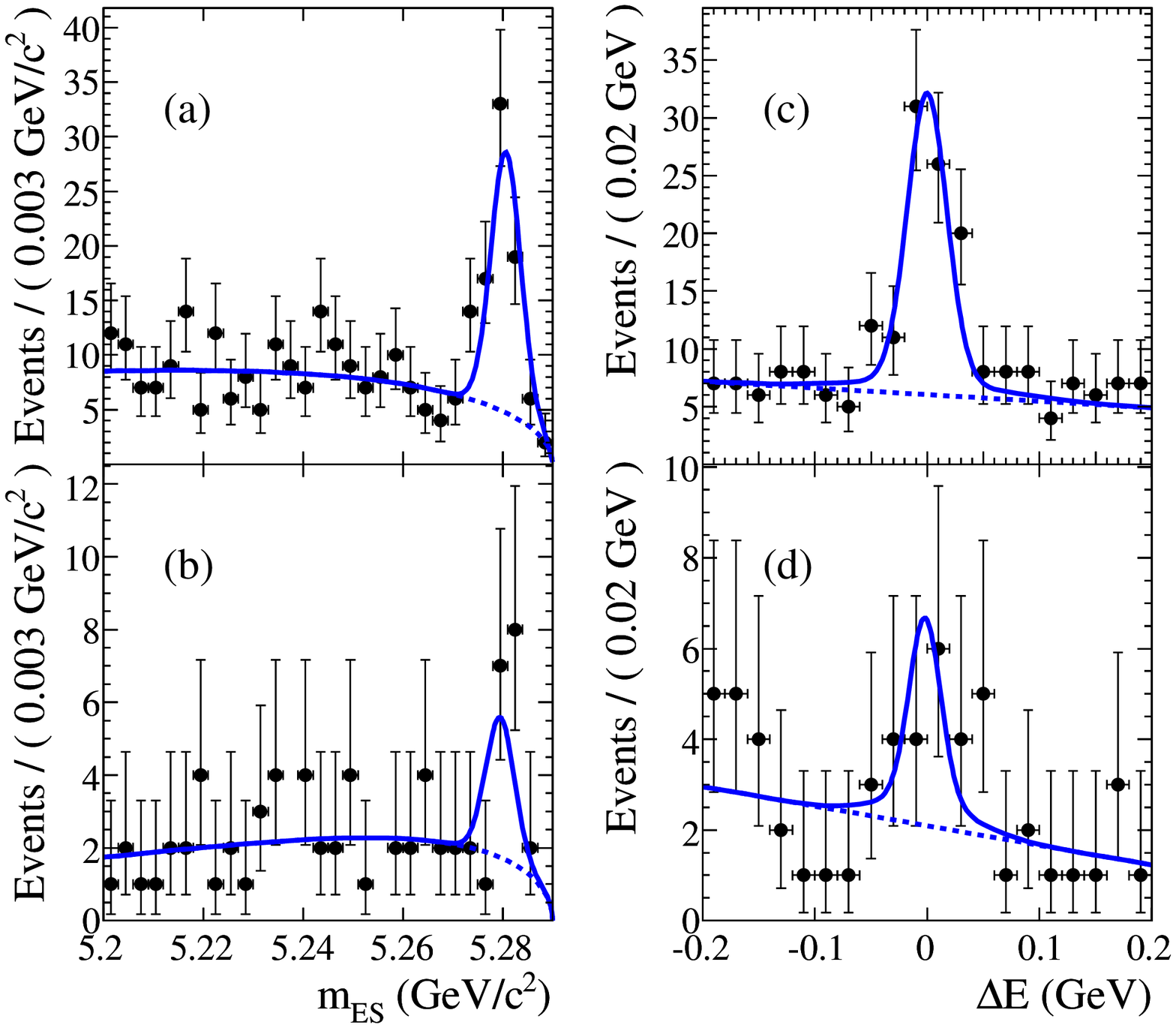}
\caption{ The projected \mes\ distributions of events with $|\de|<$ 0.05 \gev\ for (a) \btppk and (b) \bztppkz, and  the projected \de\ distributions of events with  $\mes>$ 5.27 \gevcc for (c) \btppk and (d) \bztppkz. Points with error bars represent the data, solid lines the total PDF, and dashed lines the background PDF. }
\label{data_plot}
\end{center}
\end{figure}

\begin{figure}[t]
\begin{center}
\includegraphics[width=3.5in]{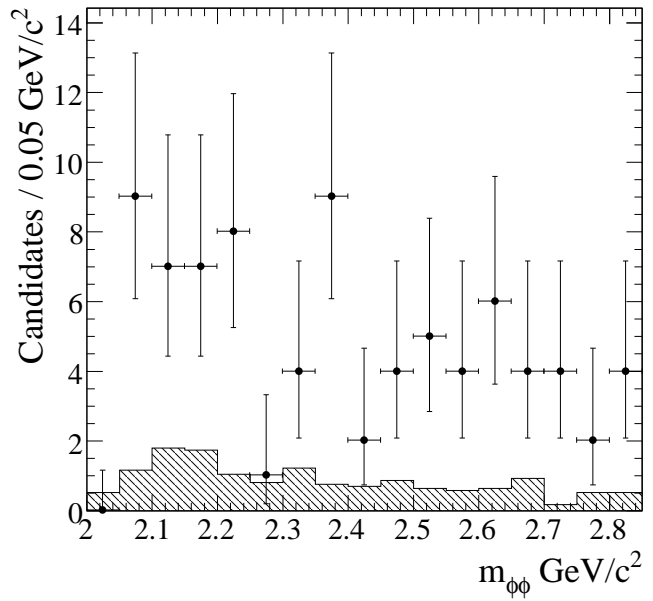}
\caption{ The projected $m_{\pp}$ distributions of events  with $|\de|<$ 0.05 \gev\  and $\mes>$ 5.27 \gevcc for the \btppk\ decay mode. The points with error bars represent the data in the signal region, and the shaded histogram represents the mass distribution of expected background from the \de\ sideband.}
\label{mpp_plot}
\end{center}
\end{figure}

  In Fig.~\ref{data_plot} (a, b), we show the \mes\ projection distributions of \btppk\ and \bztppkz\ events with a requirement $|\de|<$ 0.05 \gev. The corresponding \de\ projections for $ \mes > $ 5.27 \gevcc\ are shown in Fig.~\ref{data_plot} (c, d). The PDF model represents the data well, and a significant signal is seen in \btppk. At the present level of statistics, we do not observe any evidence for resonant structure in the \ppk\ Dalitz plot. This is consistent with our use of three-body phase space in the signal MC. The invariant mass of two \mphi\ mesons from the decay \btppk\ is shown in Fig.~\ref{mpp_plot}. Both the signal and background display smooth behavior with no evidence of any structure. We therefore see no evidence to support the hypothesis of glueball production.

  The systematic uncertainties are dominated by our knowledge of the signal and background PDFs, fit-bias correction, signal MC modeling, and possible non-resonant background contributions. The PDF-modeling error is largely included in the statistical uncertainty since most background parameters are free in the fit.~The uncertainties in the signal PDFs are estimated by varying the signal PDF parameters within their errors. We estimate the uncertainty to be 3.8$\%$ and 4.8$\%$  for charged and neutral \B\ meson decays, respectively. The systematic uncertainty due to any discrepancy in the signal PDFs between the signal MC and  the control data samples is 1.7$\%$ (1.8$\%$) for \btppk\ (\bztppkz). The uncertainty in the fit-bias correction is taken to be a half of the correction. To estimate the uncertainty due to the non-resonant background, we refit the data by including a non-resonant component in the fit. The change in the signal yield is taken as a systematic uncertainty; it is found to be 5$\%$ for the charged \B\ meson decay and $3\%$ for the neutral one. The uncertainty due to the use of three-body phase space when calculating the signal efficiency is $3\%$, as determined by the signal efficiency variation across the Dalitz plot. A correction is applied to account for known data-MC differences in track-finding efficiency. The uncertainty on this correction is $0.8\%$ per track. Systematic uncertainty due to the PID requirements are $3.5\%$ and $2.5\%$ for the charged and neutral \B\ meson decays, respectively. There is a systematic uncertainty of $2.1\%$ on the efficiency of \KS\ reconstruction. The uncertainty on the total number of \BB pairs in the data sample is $1.1\%$. Published data~\cite{PDG} provide the uncertainties in the \B-daughter product branching fractions (0.2 -- 1.4$\%$).

In conclusion, in the charged decay mode, we observe a signal of $64 \pm 9$ \stat events with a significance of 12.9 $\sigma$, corresponding to a branching fraction of \BRbtppk\ = ($7.5 \pm 1.0$ \stat $\pm$ 0.7 \syst) $\times 10^{-6}$, where $\mpp <$ 2.85 \gevcc. This result is larger than the previous measurement reported by the Belle Collaboration and is also larger than theoretical predictions. The decay \btppk\ is not dominated by a narrow glueball state with mass below 2.85 \gevcc. In the neutral mode, we observe a signal of $10.0^{+4.1}_{-3.4}$~\stat\ events with a significance of 4.2 $\sigma$, corresponding to a branching fraction of \BRbztppkz\ = ($4.1^{+1.7}_{-1.4}$ \stat $\pm$ 0.4 \syst) $\times 10^{-6}$, where $\mpp <$ 2.85 \gevcc. This is the first evidence for the process \bztppkz. The decay widths of the charged and neutral modes differ by less than $2~\sigma$. The fact that the observed charmless \mpp\ spectrum appears to extend into the region of the \etac\ resonance opens the possibility of looking for direct $\emph{CP}$ violation in interference between the two processes.

We are grateful for the excellent luminosity and machine conditions
provided by our \pep2\ colleagues, 
and for the substantial dedicated effort from
the computing organizations that support \babar.
The collaborating institutions wish to thank 
SLAC for its support and kind hospitality. 
This work is supported by
DOE
and NSF (USA),
NSERC (Canada),
IHEP (China),
CEA and
CNRS-IN2P3
(France),
BMBF and DFG
(Germany),
INFN (Italy),
FOM (The Netherlands),
NFR (Norway),
MIST (Russia), and
PPARC (United Kingdom). 
Individuals have received support from the
Marie Curie EIF (European Union) and
the A.~P.~Sloan Foundation.


\begin{thebibliography}{99}
\bibitem{Hazumi}
M. Hazumi, \plb{583}, 285 (2004).
\bibitem{F&M}
R. Fleisher and T. Mannel, \plb{511}, 240 (2001).
\bibitem{gball}
C.-K. Chua, W.-S. Hou, and S.-Y. Tsai, \plb{544}, 139 (2002).
\bibitem{Li}
Chuan-Hung Chen and Hsiang-nan Li, \jprd{70}, 054006 (2004).
\bibitem{pham}
S. Fajfer, T. N. Pham, and A. Prapotnik, \jprd{69}, 114020 (2004).
\bibitem{belle}
H.~C. Huang \etal\ [Belle Collaboration], \jprl{91}, 241802 (2003).

\bibitem{ref:babar}
B.\ Aubert \etal\ [\babar\ Collaboration], \nimBaseA~{\bf479}, 1 (2002).
\bibitem{G4}
The \babar\ detector Monte Carlo simulation is based on GEANT4: S. Agostinelli \etal, Nucl. Instr. Methods Phys. Res., Sect. A~{\bf 506}, 250 (2003).
\bibitem{fisher}
D.~M.~Asner \etal\ [CLEO Collaboration], \jprd{53}, 1039 (1996).

\bibitem{Argus}
 H.~Albrecht \etal\ [ARGUS Collaboration], \plb{241}, 278 (1990).
\bibitem{PDG}
Particle Data Group,

W.~-M.~Yao \etal, \jpg{33}, 1 (2006). 

\end{thebibliography}
\end{document}